\documentclass[preprint,review,10pt]{elsarticle}

\usepackage{amssymb}
\usepackage[hyphens]{url}
\usepackage[breaklinks=true]{hyperref}
\usepackage{pdflscape}
\usepackage{comment}

\usepackage{lineno}

\usepackage{ctable}
\usepackage{tabularx}
\newcolumntype{+}{>{\global \let \currentrowstyle \relax}}
\newcolumntype{^}{>{\currentrowstyle }}

\usepackage[cmex10]{amsmath}

\usepackage{booktabs}
\usepackage{multirow}

\newtheorem{theorem}{Theorem}[section]

\newenvironment{proof}{%
{\noindent \bf Proof. }%
}{%
\hfill$\Box$\\%
}

\usepackage{lscape}

\begin{document}

\begin{frontmatter}



\title{Effect of Vaccination to COVID-19 Disease Progression}


\author[iitmath]{Randy L. Caga-anan\corref{cor1}}
\author[upd]{Michelle N. Raza}
\author[iitmath]{Grace Shelda G. Labrador}
\author[iitmarine]{Ephrime B. Metillo}

\address[iitmath]{Department of Mathematics and Statistics, MSU-Iligan Institute of Technology, Iligan City, Philippines}

\address[upd]{Institute of Mathematics, University of the Philippines-Diliman, Quezon City, Philippines}
\address[iitmarine]{Department of Marine Science, MSU-Iligan Institute of Technology, Iligan City, Philippines}

\cortext[cor1]{Corresponding author: randy.caga-anan@g.msuiit.edu.ph}

\begin{abstract}
A mathematical model of COVID-19 with minimal compartments is developed. The model is simple enough to fit data on confirmed cases, estimate the hidden infection figure and incorporate the effect of vaccination. With the effect of the new variant being considered, the model fits well with Philippine data of confirmed cases. With the fitted parameters, results show that a second peak, strikingly similar in magnitude to the first, is to be expected if the current transmission rate is increased by about 5\% and already a complete disaster if it will go up to 20\%. The model is then used to estimate time to achieve herd immunity considering natural immunity from being infected and vaccination plan of the country. Results show that if the plan of 200,000 vaccinated people per day is achieved then herd immunity can be attained by March 2022. But if only half of that is achieved the time could go up to February 2023.
\end{abstract}


\end{frontmatter}


\section{Introduction}
The COVID-19 pandemic is a rapidly evolving health situation that immobilized nearly the entire world. It does not only affect the lives of many people but also caused much social disruptions, health issues, and havoc on the economy of many nations around the world. Coronavirus disease 2019 (COVID-19) is an infectious disease caused by severe acute respiratory syndrome coronavirus brought about by a second variant of a coronavirus type (SARS-CoV-2). This virus allegedly originated in Wuhan, China \cite{zhu}. From the epicenter of the virus, it immediately dispersed to many countries all around the world that caused a swift rise of fear and anxiety to neighboring and distant countries. On 30 January 2020, the World Health Organization declared the Chinese outbreak of COVID-19 to be a Public Health Emergency of International Concern posing a high risk to countries with vulnerable health systems \cite{sohrabi} and eventually classified it as a pandemic on 11 March 2020 \cite{WHO}.

As the COVID-19 pandemic progressed, many countries have implemented a broad range of responses. Considering the health threat brought by the deadly virus, the repercussions to the countries affected by it is to have control measures and strategies to minimize and prevent the spread of the virus. While waiting for the vaccine and its progression, countries all around the world practiced and applied non-pharmaceutical interventions(NPIs), i.e., wearing of face masks and face shields, lockdown measures, social distancing, contact tracings, quarantine measures, isolation, and public health measures and strategies. A study shows that the social distancing has a great impact for control of the outbreak \cite{hunter}. Since people who are contaminated of the virus can be symptomatic and asymptomatic, testing and isolation are necessary to stop the disease \cite{arcede2020}. Assessing the relative effectiveness of different interventions from the experience of countries to date is both challenging and crucial because many have implemented multiple (or all) of these measures with varying degrees of success. The comparison and consideration of the implications of these NPIs were looked at, focusing on the optimal level of implementation and as to what governments can do \cite{macalisang2020}. In the Philippines, local government units responded by implementing primary prevention guidelines and strict protocols to mitigate the spread of the infection. With a goal of providing science-based advice as to what the government can do to combat the rising cases of this disease, a case study of two fairly large neighboring cities in the Philippines namely Iligan City and Cagayan de Oro City was carried out to analyze which mitigating strategies applied are much effective. Results indicated that social distancing and age specific quarantine can effectively slow down the progression of the disease \cite{nuss}. Moreover, social distancing combined with an effective testing strategy can keep the epidemic at sub-critical level \cite{bock2020}. 

Since the emergence of COVID‐19, there has been an explosion of vaccine development. Consequently, while applying the best mitigating strategies, the availability of the vaccines was anticipated and its eventual use looked forward by countries. Vaccines by definition are biological agents that elicit an immune response to a specific antigen derived from an infectious disease-causing pathogen. The first vaccine was developed in 1796 and since then, vaccines have helped to suppress the spread of several infectious diseases. A range of vaccine development approaches for SARS-CoV-2 has been proposed and being tested in clinical trials. These include traditional approaches – inactivated, live attenuated and protein/adjuvant methods, and more novel, as yet, unlicensed techniques  – viral vectors and nucleic acids. By 24 September 2020, the SARS‐CoV‐2 vaccine landscape included 43 candidates being tested in clinical trials and more than 200 candidates are awaiting trial \cite{mellet}. Apparently, different vaccine platforms have advantages and disadvantages. Several factors need to be considered before any vaccine progresses to widespread usage. First and foremost is vaccine safety and efficacy. Closely linked is the issue on the scope for global scale‐up manufacture to produce enough doses essential to achieving herd immunity \cite{treg}. 

After the announcement of COVID-19 vaccine efficacy through clinical trials by several manufacturers, a comprehensive post-efficacy strategy for the next steps to ensure vaccination of the global population is now required. These considerations should include how to manufacture billions of doses of high-quality vaccines, support for vaccine purchase, coordination of supply, the equitable distribution of vaccines and the logistics of global vaccine delivery, all of which are a prelude to a massive vaccination campaign targeting people of all ages \cite{kim}. Countries all around the world began their vaccination efforts and campaign. The Philippines, specifically, launched its national coronavirus vaccination campaign on 1 March, 2020 amid widespread public skepticism and a struggle to procure vaccines. The Philippines is the last Southeast Asian nation to receive a COVID-19 vaccine supply \cite{voa}. Finally, the beginning of the vaccinations all around the world gives a glimpse of hope to a future where the virus will be kept at bay, if not, totally eradicated. Still, the level of vaccination efforts takes significance to the urgency of the herd immunity to be achieved.

In this paper, we analyze COVID 19 progression with vaccination and provide an answer on how long will it take to achieve herd immunity. The aim of this study is to incorporate the effect of vaccination and estimate time to achieve herd immunity considering natural immunity from being infected and the vaccination plan of the country and to what levels of vaccination the government can do to achieve the targeted time.

\section{Mathematical Model}
Since data are scarce we created a simple minimalist model of COVID-19 disease progression that could easily fit the publicly available data of confirmed cases. Generally, before being confirmed positive by testing, an individual starts being a susceptible prior to getting infected. After some time, symptoms will emerge and the individual may either get hospitalized or noticed by the authorities, and so they will be tested. After a period of time, the individual will likely recover but in some cases will die. We use the term removed to mean either of the cases. 

\begin{figure}[htbp]
	\centering
	\includegraphics[scale=0.5]{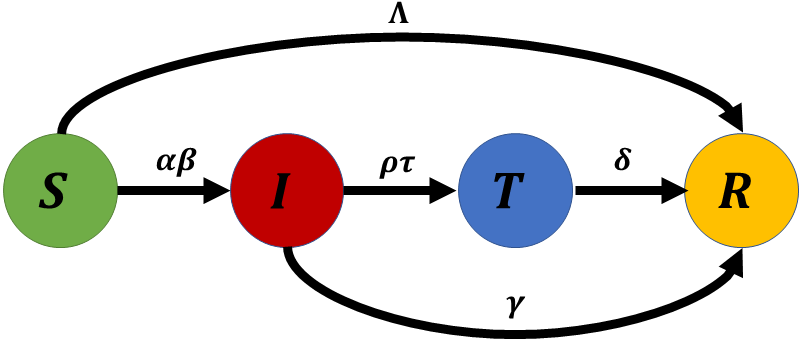} 
	\\[-0.3cm]\caption{Simulation fit with data}
	\label{fig1}
\end{figure}

Thus, in our model, the population is classified into four compartments: Susceptible $(S)$, Infected $(I)$, Tested positive $(T)$, and Removed $(R)$. Over the time span of our simulations, we note that the number of births and deaths in the country does not significantly affect the size of the total population. Hence, we  adapted a closed population model and we denote by $N_0$ the total population. We note that the compartments are non-negative for all time.

We assume that those vaccinated and those who are already infected will have developed immunity long enough until herd immunity and so we do not include reinfection on our model.

The model is governed by the following system of ordinary differential equations:
\begin{align}
\frac{d S}{d t} &=-\alpha\beta_0 SI-\Lambda_S \label{eq1}\\
\frac{d I}{d t} &= \alpha\beta_0 SI- (\gamma + \rho\tau)I \label{eq2}\\
\frac{d T}{d t} &= \rho\tau I - \delta T \label{eq3}\\
\frac{d R}{d t} &= \gamma I + \delta T + \Lambda_S \label{eq4}
\end{align}
where $\beta_0=\beta/N_0$ and $\Lambda_S=\min\{\Lambda,S\}$.
We use the minimum operator to ensure that the vaccination rate is at most equal to the number of susceptible.
The parameters are described in Table~\ref{tab:parameters}.
\begin{table}[htbp]
	\centering
\begin{tabular}{c|c|c}
	\hline 
Parameter	& Description & Unit \\ 
	\hline 
	$\Lambda$ & vaccination rate &    persons/day \\
	\hline 
	$\beta$ & transmission rate  &   1/day \\ \hline 
	$\alpha$ & transmission reduction control&   dimensionless \\
	\hline 
	$\rho$ &  proportion of infection detected &  dimensionless \\ 
	\hline 
	$\tau$ & average detection time  &  1/day  \\ 
	\hline
	$\delta$ &  removal rate from $T$ to $R$  & 1/day  \\ 
	\hline 
	$\gamma$ &  removal rate from $I$ to $R$  & 1/day  \\ 
	\hline 
\end{tabular} 
\caption{Parameters of the model}
\label{tab:parameters}
\end{table}

\section{Analysis}
Equating equations (\ref{eq1})-(\ref{eq4}) to zero and then adding the resulting equations from (\ref{eq1}) and (\ref{eq2}) (or (\ref{eq3}) and (\ref{eq4})), we get
\begin{equation*}
    I=\frac{-\Lambda_S}{\rho\tau + \gamma}=\frac{-\min\{\Lambda,S\}}{\rho\tau + \gamma}.
\end{equation*}
Since $\rho\tau + \gamma > 0$ and $I\geq 0$, either $\Lambda = 0$ or $S=0$. Hence, we have the unique equilibrium point $(0,0,0,N_0)$ if $\Lambda > 0$ or $(S^*,0,0,R^*)$, where $S^*=N_0-R^*$, if $\Lambda = 0$.

\subsection{Basic Reproduction Number}
By definition, $\mathcal{R}_0$ denotes the average number of individuals directly infected by a single infected individual over the duration of its infectious period in a population without any deliberate intervention to stop its spread. Hence, we will compute $R_0$ only for the case when there is no vaccination yet or $\Lambda = 0$ and the unique equilibrium point is $(S^*,0,0,R^*)$, where $S^*=N_0-R^*$.

We will compute for $\mathcal{R}_0$ using the next generation operator approach defined by Diekman et. al \cite{diek} and Driessche and Watmough \cite{dri}.
It is standard to check that the domain
\begin{equation*}
    \Omega = \left\{({\mathit{S},\mathit{I}, \mathit{T},\mathit{R}) \in \mathbb{R}_{+}^{4} ; \ 0 \leqslant \mathit{S},\mathit{I}, \mathit{T},\mathit{R} \leqslant N(0)}\right\}
\end{equation*}
is positively invariant. In particular, there exists a unique global in time solution ($\mathit{S}, \mathit{I}, \mathit{T}, \mathit{R})$ in  $\mathit{C}(\mathbb{R}_{+} ; \Omega $) as soon as the initial condition lives in $\Omega $. 
Since the infected individuals are in $\mathit{I}$ and $\mathit{T}$, the rate of new infections in each compartment $(\mathcal{F})$ and the rate of other transitions between compartments $(\mathcal{V})$ can be rewritten as
\begin{center} 
\[
\mathcal{F} = \begin{pmatrix} 
\alpha\beta_0 SI \\
0
\end{pmatrix} , \ \ \ \ \ \
\mathcal{V} = \begin{pmatrix} 
(\gamma + \rho\tau)I \\
-\rho\tau I + \delta T
\end{pmatrix} 
\] \\
\end{center}
Thus,
\begin{center} 
\[
F = 
\begin{pmatrix} 
\alpha\beta_0 S & 0  \\
0 & 0
\end{pmatrix}
\]
\end{center}
and
\begin{center}

$$V =
\begin{pmatrix} 
\gamma + \rho\tau & 0 \\
-\rho\tau  & \delta
\end{pmatrix} ,
 \qquad
V^{-1} = 
\begin{pmatrix}
\frac{1}{\gamma + \rho\tau} & 0 \\
\frac{\rho\tau}{(\gamma + \rho\tau)\delta} & \frac{1}{\delta}
\end{pmatrix} $$
\end{center} 
Therefore, the next generation matrix is 
\begin{center}
\[
FV^{-1} = 
\begin{pmatrix}
\frac{\alpha \beta_0 S}{\gamma + \rho\tau}  & 0 \\
0 & 0
\end{pmatrix}
\]
\end{center}
where $\beta_0=\beta/N_0$. \\
Hence, the basic reproduction number $\mathcal{R}_0$ \  for \ the \ Disease \ Free \ Equilibrium \ ($\textit{S}^*$, 0, 0, $\textit{R}^*$), with \ $\textit{N}^*$ = $\textit{S}^*$ + $\textit{R}^*$, \ is \
$$\mathcal{R}_0 = \frac{\alpha \beta_0}{\gamma + \rho\tau}\mathit{S^{*}}.$$

\subsection{Stability of the unique equilibrium point}
By computing the eigenvalues of the Jacobian matrix, we can deduce that if $R_0 < 1$, then the disease-free equilibrium point is locally asymptotically stable. Moreover, the next theorem will show that the asymptotic behavior does not depend on $R_0$. That is, for all initial data in $\Omega$, the
solution converges to the disease-free equilibrium point when time goes to infinity. 
\begin{theorem}
The DFE $(S^*,0,0,R^*)$ or $(0,0,0,N_0)$ is the unique positive equilibrium and it is globally asymptotically stable.
\end{theorem}

\proof
From the last differential equation in our system, we can deduce that ${R}$ is an increasing function bounded by ${N(0)}$. Thus ${R(t)}$ converges to ${R}^{*}$ as ${t}$ goes to $+\infty$. Integrating over time we get
$$
\textit{R(t) - R(0)} = \int_0^t  \gamma I + \delta T + \Lambda_S
$$
and
$$
\textit{R* - R(0)} = \int_0^{+\infty} \gamma I + \delta T + \Lambda_S,$$
which is finite. Furthermore, $\gamma I + \delta T + \Lambda_S$ goes to 0 as $\textit{t} \xrightarrow \ +\infty$, and each term of this sum does thanks to the positivity of the solution. Adding the two first equations implies that
$$(S + I)' = -\Lambda_S - (\gamma + \rho\tau)I$$
and S + I is a nonnegative decreasing function whose derivative tends to zero. Then, $I(t) \rightarrow_{t\rightarrow +\infty} 0$ and $S(t) \rightarrow_{t\rightarrow+\infty} S^{*}$.
\qed \\

\section{Simulations}

\subsection{Parameter values}
From \cite{lauer2020}, latency and infection periods have been estimated as 5 and 7 days, respectively. We assume that most of the people tested are showing at least mild symptoms already, and so we use  $\tau = 1/5$. Noting of the infection period, we use $\delta = 1/7$. We add 7 more days to those who tested positive before we label them removed and so we use $\gamma = 1/7$.

For the parameters $\beta, \alpha$ and $\rho$, we estimated its values by fitting the model with the cumulative confirmed cases of the Philippines from January 30, 2020 to February 27, 2021. The data set is being curated by the Johns Hopkins University \cite{dong2020}. We acknowledge that in reality the three parameters are varying through time due to the varying controls being implemented and the evolution of the virus. Thus, in our parameter estimation we let $\beta, \alpha$ and $\rho$ to be piecewise functions as follows:
\begin{equation}
\beta =
    \begin{cases}
    \beta_1, \text{ for $0\leq t < 343$ (Jan. 7, 2021)}\\
    \beta_1(1+\beta_2), \text{ for $t\geq 343$,}
    \end{cases}
\end{equation}
where $\beta_2$ corresponds to the increase of the transmission rate due to the introduction of the more transmissible new variant of the virus in the population,
\begin{equation}
\alpha =
    \begin{cases}
    \alpha_1, \text{ for $0\leq t < 46$ (March 16, 2020)}\\
    \alpha_2, \text{ for $46\leq t < 123$ (June 1, 2020)}\\
    \alpha_3, \text{ for $123\leq t < 320$ (Dec. 15, 2020)}\\
    \alpha_4, \text{ for $320\leq t < 340$ (Jan. 4, 2021)}\\
    \alpha_5, \text{ for $340\leq t < 396$ (March 1, 2021)}\\
    \alpha_6, \text{ for $t\geq 396$,}
    \end{cases}
\end{equation}
where the dates correspond to the noticeable changes in the control measures of the country and we put in $\alpha_6$ our estimate for the future transmission reduction control,
\begin{equation}
\rho =
    \begin{cases}
    \rho_1, \text{ for $0\leq t < 184$ (Aug. 1, 2020)}\\
    \rho_2, \text{ for $t\geq 343$,}
    \end{cases}
\end{equation}
where $\rho_2$ corresponds to the start of the current testing rate of the country.

The obtained parameter values are given in Table~\ref{tab:parameters2} and the fit is shown in Figure~\ref{fig1}.
\begin{table}[htbp]
	\centering
\begin{tabular}{c|c|c|c|c|c}
	\hline 
$\alpha$ & Value & $\beta$ & Value & $\rho$ & Value \\ 
	\hline 
$\alpha_1$ &  1  & $\beta_1$ & 0.3713 & $\rho_1$ & .10814\\
	\hline 
$\alpha_2$ &  0.44689  & $\beta_2$ & $1.2658\times 10^{-13}$ & $\rho_2$ & 0.30196\\
	\hline
$\alpha_3$ &  0.52219  &  & &  & \\
	\hline
$\alpha_4$ &  0.6516  &  & &  & \\
	\hline
$\alpha_5$ &  0.55152  &  & &  & \\
	\hline
\end{tabular} 
\caption{Fitted parameter values}
\label{tab:parameters2}
\end{table}

\begin{figure}[htbp]
\centering
\includegraphics[scale=0.4]{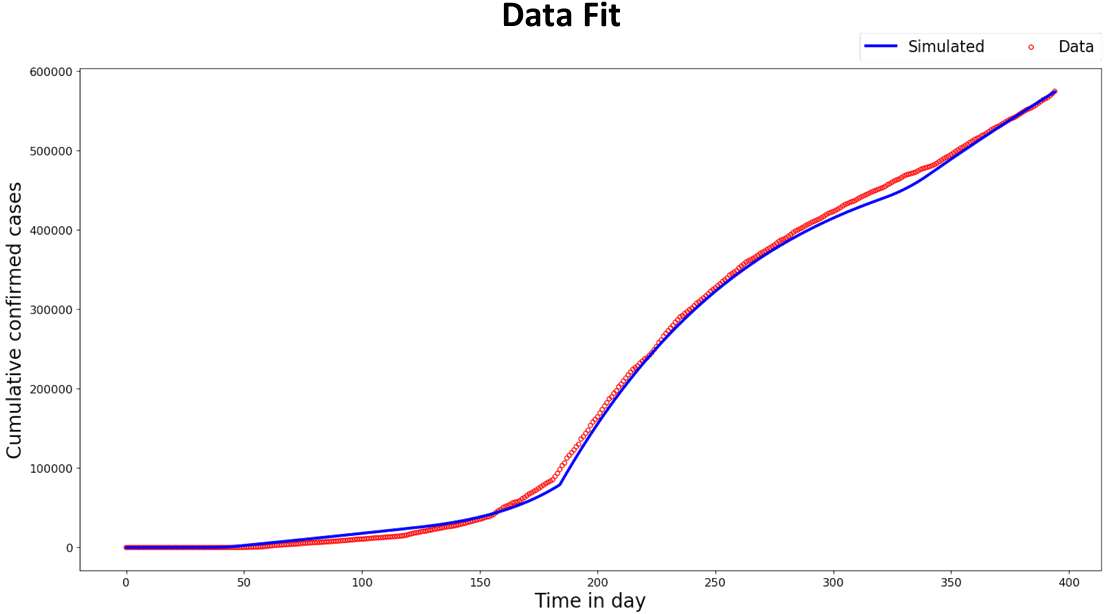} 
\\[-0.5cm]\caption{Simulation fit with data}
\label{fig1}
\end{figure}

For the vaccination parameter, we note that the country is targeting to have 200,000 people vaccinated per day. However, looking at Indonesia’s vaccination data, that rate could be achieved only gradually.  So, in our simulations, we estimate 10,000 vaccinations daily for the first 10 days, then 50,000 daily for the next 15 days. Afterwards, we increase it to 100,000 daily for another 15 days and then 150,000 daily for yet another 15 days, and then settle at 200,000 per day after that. 

\subsection{Without vaccination}
\begin{figure}[htbp]
\centering
\includegraphics[scale=0.4]{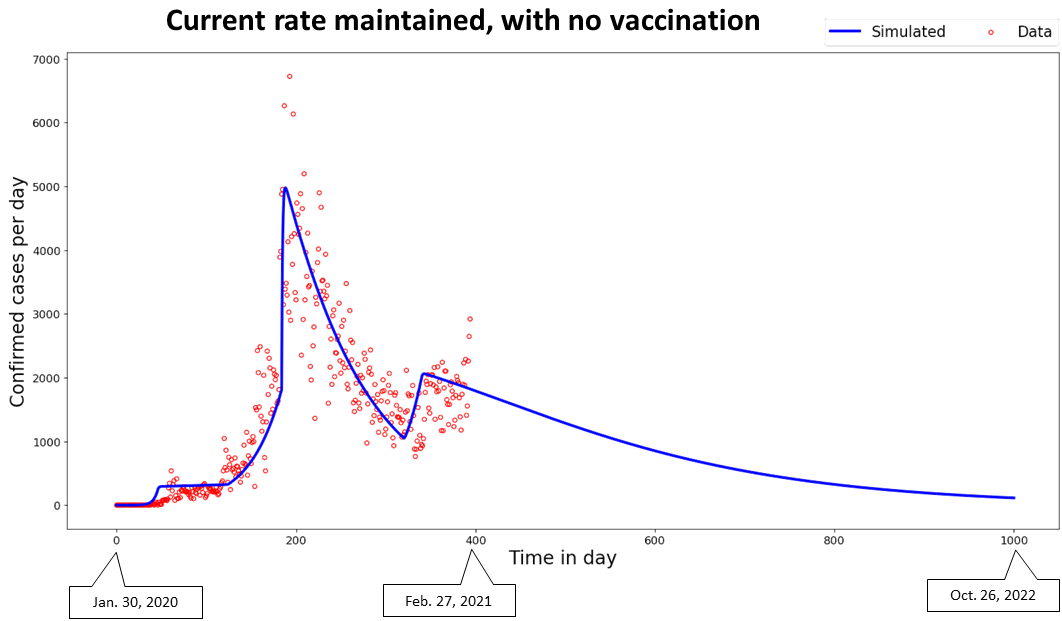} 
\\[-0.5cm]\caption{No increase from the current fitted transmission rate}
\label{fig2}
\end{figure}

\begin{figure}[htbp]
\centering
\includegraphics[scale=0.4]{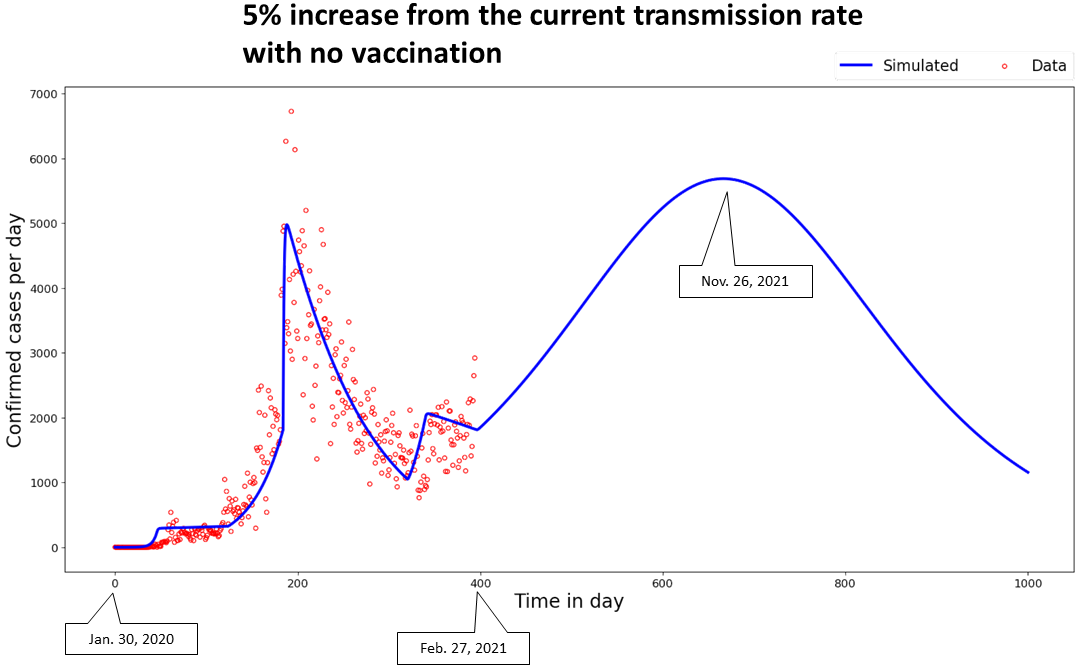} 
\\[-0.5cm]\caption{5\% increase from the current fitted transmission rate}
\label{fig3}
\end{figure}

\begin{figure}[htbp]
\centering
\includegraphics[scale=0.4]{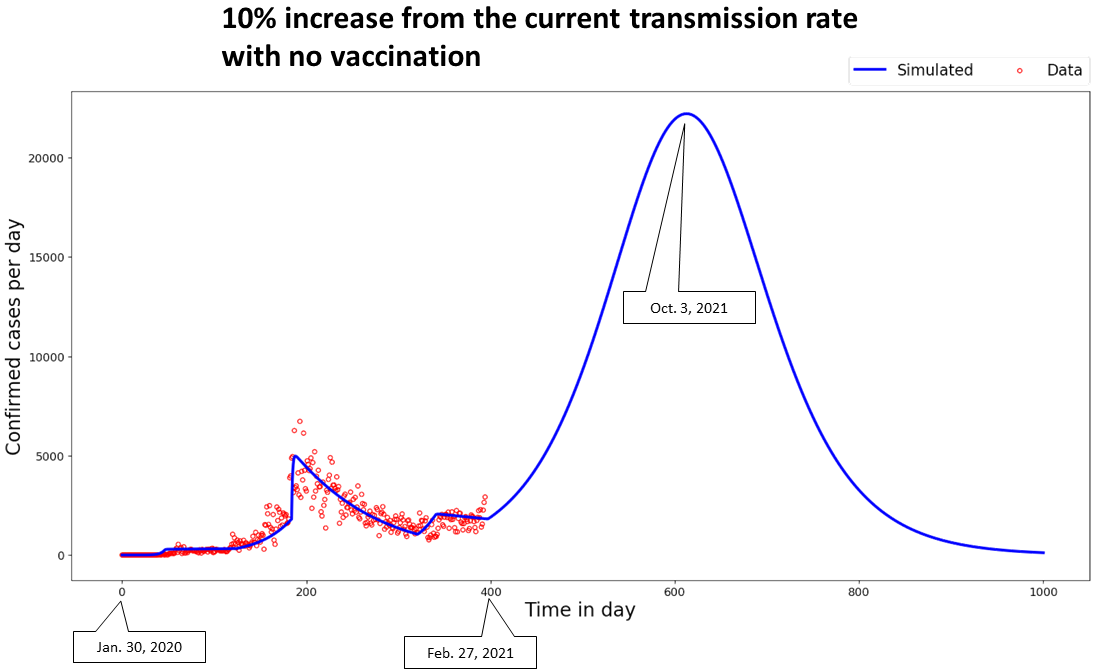} 
\\[-0.5cm]\caption{10\% increase from the current fitted transmission rate}
\label{fig4}
\end{figure}

\begin{figure}[htbp]
\centering
\includegraphics[scale=0.4]{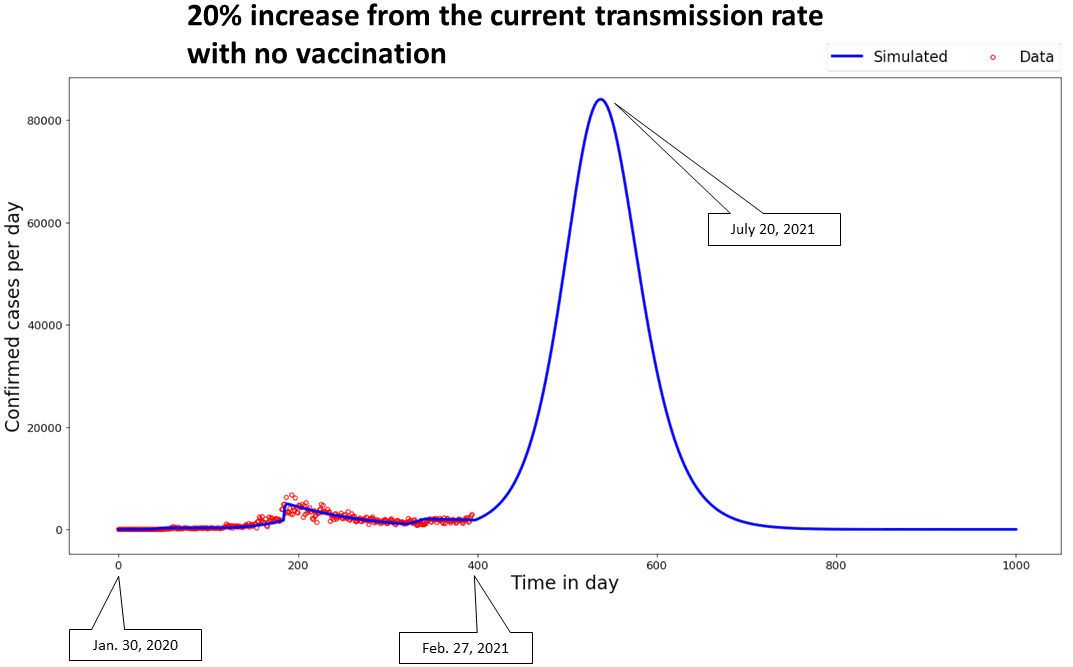} 
\\[-0.5cm]\caption{20\% increase from the current fitted transmission rate}
\label{fig5}
\end{figure}

We consider the case without vaccination to see the likely progression of the disease if we vary the value of $\alpha_6$, considering that some travel restrictions are being lifted in the country. The simulation results are given by Figures \ref{fig2}, \ref{fig3}, \ref{fig4}, and \ref{fig5}.

\subsection{With vaccination}
\begin{figure}[htbp]
\centering
\includegraphics[scale=0.4]{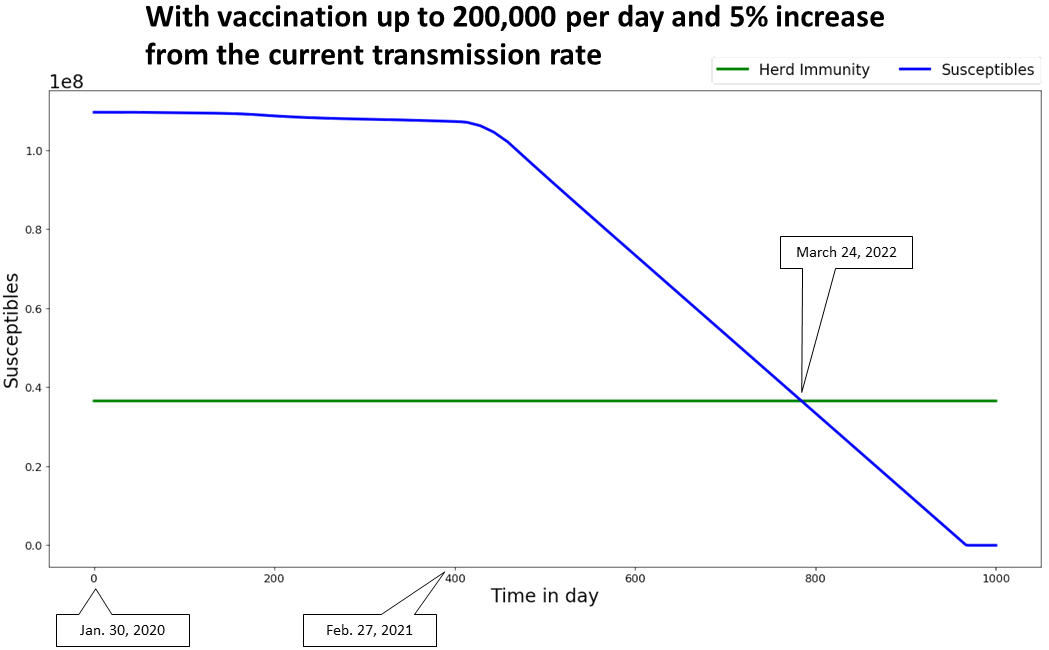} 
\\[-0.5cm]\caption{Herd immunity in the case where maximum vaccinated per day is 200000 with 5\% increase from the current fitted transmission rate}
\label{fig6}
\end{figure}

\begin{figure}[htbp]
\centering
\includegraphics[scale=0.4]{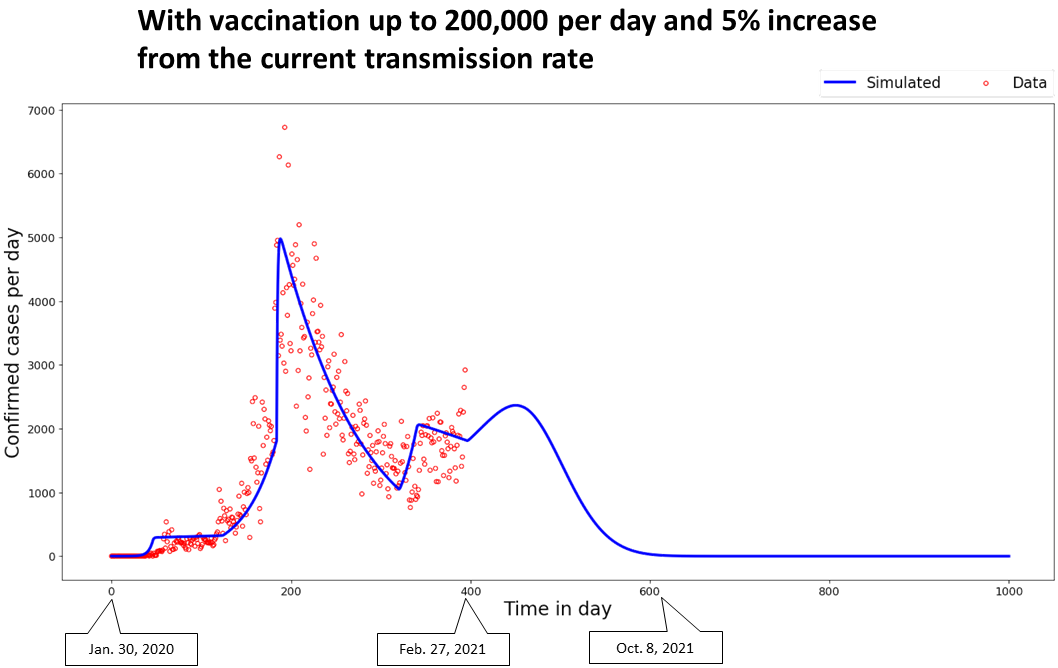} 
\\[-0.5cm]\caption{Per day confirmed cases in the case with maximum vaccinated per day is 200000 with 5\% increase from the current fitted transmission rate}
\label{fig7}
\end{figure}

\begin{figure}[htbp]
\centering
\includegraphics[scale=0.4]{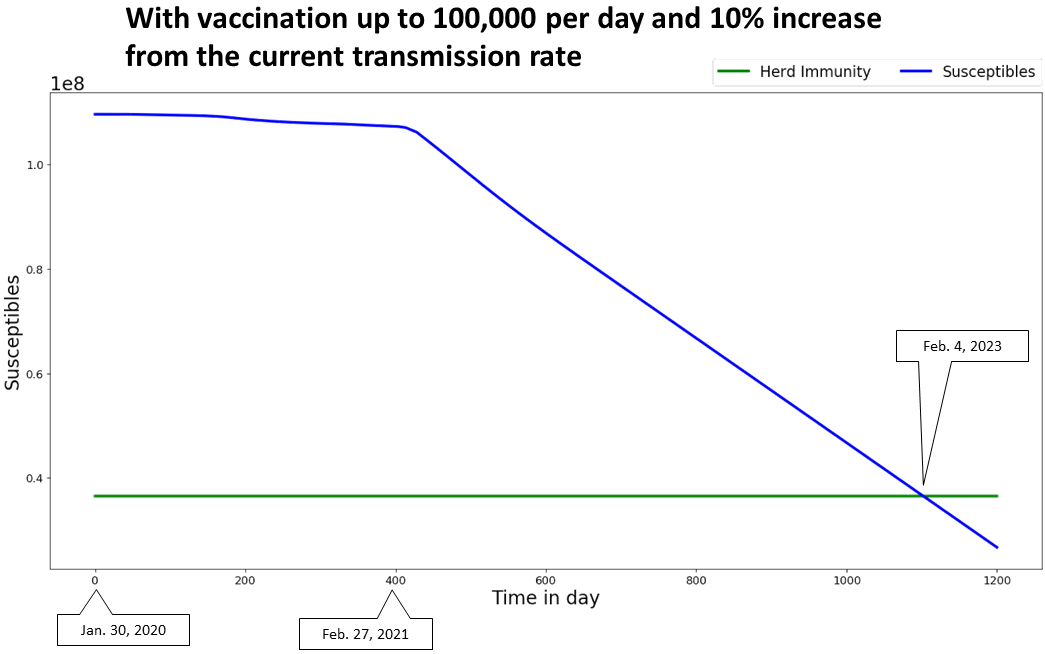} 
\\[-0.5cm]\caption{Herd immunity in the case where maximum vaccinated per day is 100000 with 10\% increase from the current fitted transmission rate}
\label{fig8}
\end{figure}

\begin{figure}[htbp]
\centering
\includegraphics[scale=0.4]{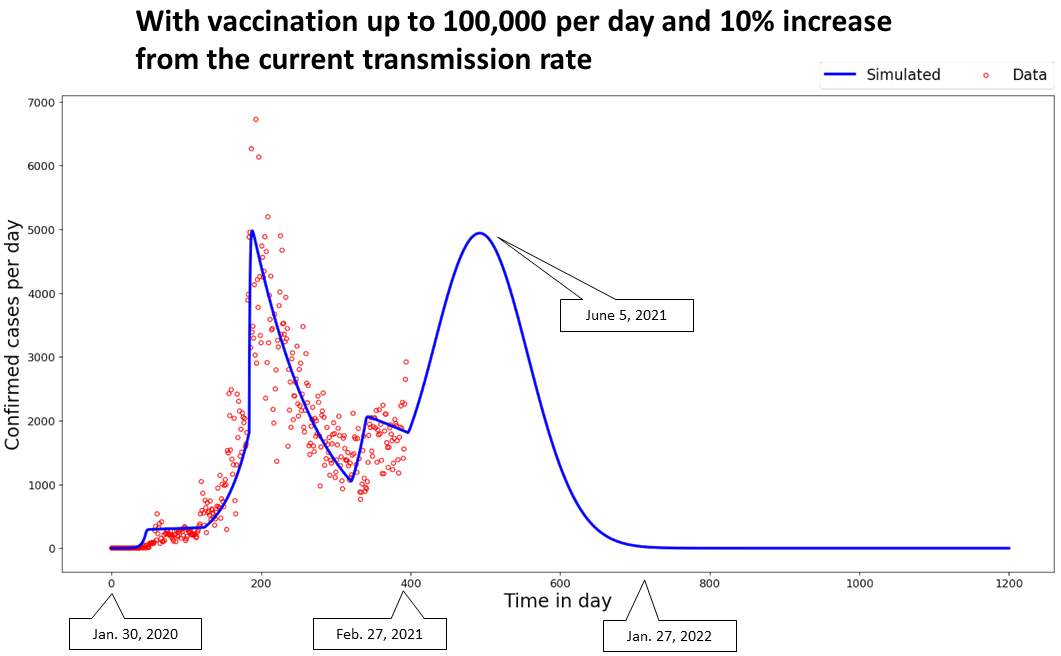} 
\\[-0.5cm]\caption{Per day confirmed cases in the case with maximum vaccinated per day is 100000 with 10\% increase from the current fitted transmission rate}
\label{fig9}
\end{figure}

In the case when there is already a vaccination campaign, we want to know when will the country achieved herd immunity factoring also the natural immunity brought by the infected class. We also would like to know the disease progression with the vaccination campaign. The results are given in Figures \ref{fig6}, \ref{fig7}, \ref{fig8}, and \ref{fig9}.

\section{Discussion}
The effect of the new variant, that is increasing the transmission rate, is incorporated in the model around the early part of January 2021, during the time when the news that the new strain has entered the country was released. It is noteworthy that fitting the model with data on the new variant being incorporated in the transmission rate around the time it was detected in the country, produced a very good result, far better than without it, suggesting that the country is already experiencing the effect of the new variant.

The results here are for best case scenarios and so the time for herd immunity given the rate in the simulations is the earliest possible. If we factor in reinfection and vaccine efficacy the the herd immunity threshold could go up and the time to reach herd immunity will be longer.

\subsection{Scenarios without vaccination}

Considering that some travel restrictions are already being loosened, we found it interesting to simulate the spread of the disease, with varying increase in the fitted transmission reduction control. 

When we increased by 5\% the current fitted transmission reduction control ($\alpha_5$ to $\alpha_6$, which may intuitively mean collectively “lowering our guard” by 5\%), the model showed that we are to expect a second “peak”, very similar in magnitude to the first peak, around November 26, 2021. “Lowering our guard” by 10\%, or increasing the current fitted transmission rate by 10\%, yielded a result of having a second “peak” of about 22,000 cases in a day by around October 3, 2021. And “lowering” it further to 20\% is a complete disaster, with the second “peak” to occur around July 20, 2021 with about 84,000 cases in a day.

It is very important to note that an increase in mobility does not necessarily mean an increase in transmission rate. For instance, even if everyone is traveling all the time but everyone is also wearing a complete personal protective equipment, or in an extreme, but ideal case, a full body bio hazard suit, then we will not expect an increase in transmission rate and we may even expect a decrease instead.

So, it pays very much to never let our guard down with face mask and face shield always on when in public or in risky areas. Figure \ref{fig2} shows that the country can avoid another exponential increase in cases.

\subsection{Scenarios with vaccination}

On March 1, 2021, the country started its vaccination campaign and so we wish to know when can the country attain herd immunity (when 2/3 of the country’s population is already immune) as a result of the combination of natural immunity from being infected and the country’s targeted 200,000 vaccinations per day. If we look at Indonesia’s vaccination data, they have done it gradually at first. So, in our simulations, we estimate 10,000 vaccinations daily for the first 10 days, then 50,000 daily for the next 15 days. Afterwards, we increase it to 100,000 daily for another 15 days and then 150,000 daily for yet another 15 days, and then settle at 200,000 per day after that. Assuming the increase in “transmission” rate is around 5\% to 10\%, the vaccination has a very good effect at avoiding a disastrous second “peak”, with herd immunity attained around the end of March 2022. Moreover, around October 2021, confirmed cases could be less than 10 individuals per day.

If only 100,000 vaccinations per day is achieved, it could still help avoid a disastrous second “peak” but herd immunity will be attained around February 2023.


\section{Acknowledgement}
The authors would like to thank DOST and MSU-IIT of the Philippines.


\bibliographystyle{plain}



\end{document}